%% file: main.tex
\documentclass{sig-alternate-10pt}

\usepackage{lastpage,framed,url,caption,subcaption,graphicx,flushend,cleveref}
\usepackage{amsmath,xcolor,soul,algpseudocode,algorithm,xspace,tabularx,multirow}
\usepackage{amssymb}

\PassOptionsToPackage{hyphens}{url}

\expandafter\def\expandafter\UrlBreaks\expandafter{\UrlBreaks
  \do\a\do\b\do\c\do\d\do\e\do\f\do\g\do\h\do\i\do\j%
  \do\k\do\l\do\m\do\n\do\o\do\p\do\q\do\r\do\s\do\t%
  \do\u\do\v\do\w\do\x\do\y\do\z\do\A\do\B\do\C\do\D%
  \do\E\do\F\do\G\do\H\do\I\do\J\do\K\do\L\do\M\do\N%
  \do\O\do\P\do\Q\do\R\do\S\do\T\do\U\do\V\do\W\do\X%
  \do\Y\do\Z}

\newcount\Comments  \Comments=0   
\newcommand{\kibitz}[2]{\ifnum\Comments=1\textcolor{#1}{#2}\fi}
\newcommand{\rish}[1]  {\kibitz{red}   {[\textbf{Rishab} -- \emph{#1}]}}
\newcommand{\Pnote}[1]  {\kibitz{purple}   {[\textbf{PG} -- \emph{#1}]}}

\providecommand{\vs}{vs. }
\providecommand{\ie}{\emph{i.e.,} }
\providecommand{\eg}{\emph{e.g.,} }

\providecommand{\etal}{\emph{et al.\xspace}}
\providecommand{\etc}{\emph{etc.\xspace}}
\providecommand{\via}{\emph{via} }

\providecommand{\systemname}{Cipollino\xspace}

\providecommand{\myparab}[1]{\smallskip\noindent\textbf{#1} }

\newenvironment{packeditemize}{\begin{list}{$\bullet$}{\setlength{\itemsep}{0.2pt}\addtolength{\labelwidth}{0pt}\setlength{\leftmargin}{\labelwidth}\setlength{\listparindent}{\parindent}\setlength{\parsep}{1pt}\setlength{\topsep}{0pt}}}{\end{list}}
\newcommand{\squishenum}{   \begin{enumerate}{}    { \setlength{\itemsep}{0pt}      \setlength{\parsep}{0pt}      \setlength{\topsep}{3pt}       \setlength{\partopsep}{0pt}      \setlength{\leftmargin}{1.5em} \setlength{\labelwidth}{1em}      \setlength{\labelsep}{0.5em} } }
\newcommand{\squishlist}{   \begin{list}{$\bullet$}    { \setlength{\itemsep}{0pt}      \setlength{\parsep}{3pt}      \setlength{\topsep}{3pt}       \setlength{\partopsep}{0pt}      \setlength{\leftmargin}{1.5em} \setlength{\labelwidth}{1em}      \setlength{\labelsep}{0.5em} } }
\newcommand{\squishlisttwo}{   \begin{list}{$\bullet$}    { \setlength{\itemsep}{0pt}    \setlength{\parsep}{0pt}      \setlength{\topsep}{0pt}     \setlength{\partopsep}{0pt}      \setlength{\leftmargin}{2em} \setlength{\labelwidth}{1.5em}      \setlength{\labelsep}{0.5em} } }
\newcommand{\squishend}{    \end{list}  }
\newcommand{\squishenumend}{	\end{enumerate}	}

\begin{document}

\date{}

\title{Holding all the ASes: Identifying and Circumventing the Pitfalls of
AS-aware Tor Client Design}

\numberofauthors{1}
\author{
\alignauthor 
Rishab Nithyanand \hfill Rachee Singh \hfill Shinyoung Cho \hfill Phillipa Gill\\ 
\affaddr{Stony Brook University}\\
\normalsize{\{rnithyanand, racsingh, shicho, phillipa\}@cs.stonybrook.edu}
}

\maketitle


\begin{abstract}

Traffic correlation attacks to de-anonymize Tor users are possible when an
adversary is in a position to observe traffic entering and exiting the Tor
network. Recent work has brought attention to the threat of these attacks by
network-level adversaries (\eg Autonomous Systems). We perform a historical
analysis to understand how the threat from AS-level traffic correlation attacks
has evolved over the past five years. We find that despite a large number of
new relays added to the Tor network, the threat has grown. This points to the
importance of increasing AS-level diversity in addition to capacity of the Tor
network.

We identify and elaborate on common pitfalls of AS-aware Tor client design and
construction. We find that succumbing to these pitfalls can negatively impact
three major aspects of an AS-aware Tor client -- (1) security against  AS-level
adversaries, (2) security against relay-level adversaries, and (3) performance.
Finally, we propose and evaluate a Tor client -- \systemname -- which avoids
these pitfalls using state-of-the-art in network-measurement. Our evaluation
shows that \systemname is able to achieve better security against network-level
adversaries while maintaining security against relay-level adversaries and
performance characteristics comparable to the current Tor client.

\end{abstract}

\input{introduction}

\input{background}

\input{measurement}

\input{challenges}

\input{cipollino}

\input{conclusions}

\myparab{Data and source-code release:} In an effort to enable reproducibility
and ease future comparative evaluation efforts, the following resources will be
made available on acceptance of this work: the \systemname Tor client, the
destination-based graphs provided by PathCache during the time of this study,
and the Web and mixed-application user-models used in our simulations.

{\bibliographystyle{unsrt}
\bibliography{./bibliography}}

\end{document}

%% file: introduction.tex
\section{Introduction}\label{sec:introduction}

As governments and organizations increase their commitment to mass surveillance
and online tracking, the Tor anonymity network has become the de facto
technology for preserving anonymity and privacy on the Internet with nearly two
million daily users \cite{tor-metrics}.

Tor's popularity has made it a prime target for attacks and also increases the
importance of improving its defenses. In this paper, we focus on a long-standing
class of attacks known as traffic-correlation attacks. In a traffic-correlation
attack, an adversary correlates the characteristics of traffic (\eg packet
sizes, inter-packet timings, \etc) entering and exiting the Tor network.
Successfully correlating these flows results in the de-anonymization of Tor
users -- \ie it becomes possible to identify the destination server being
contacted by a Tor user.

Traffic correlation attacks have been known about for over a decade
\cite{Feamster-WPES04} but as our study shows, Tor is still incredibly
vulnerable. Worse yet, the recent Snowden leaks have confirmed that the NSA and
GCHQ, in collusion with several Internet Service Providers (ISPs), have actively
been working to implement network-level attacks in the wild
\cite{Schneier-NSA1, Guardian-NSA2, Guardian-NSA3}. In order to launch a traffic
correlation attack, an adversary needs to be able to observe network traffic on
(1) the path between the Tor user and the entry (relay) to the Tor network and
(2) the path between the exit (relay) from the Tor network to the destination
server. Such attacks have been shown to be feasible for both, relay-level
adversaries \cite{Cai-CCS12, Wang-WPES13, Wang-Security14} and network-level
adversaries such as Autonomous Systems (ASes) \cite{Feamster-WPES04,
Murdoch-SP05, Murdoch-PETS07, Johnson-CCS13, Juen-PETS15, Sun-Security15}. 
 
While the problem of relay-level traffic-correlation attacks have been mitigated
and solutions have been integrated into the current Tor client \cite{Cai-CCS14,
Dingledine-PETS14}, the problem of defending against network-level
traffic-correlation attacks remains unsolved. While numerous defenses have been
proposed \cite{Feamster-WPES04, Edman-CCS09, Akhoondi-SP12, Nithyanand-NDSS16},
none have been successfully adopted in practice. We identify five pitfalls that
render existing AS-aware Tor clients insecure, impractical, or both. We
characterize the impact of these pitfalls and propose a modified Tor client that
is able to mitigate them. 



In this paper we make three major contributions.

\myparab{Measuring the threat (Section~\ref{sec:measurement}).} We perform a
current and a historical analysis to understand how the threat from AS-level
adversaries has evolved over the past five years. From these measurements, we
make the following observations:

\begin{itemize}
\item When considering Tor clients used specifically for the purpose of loading
webpages, 31\% of the circuits constructed by the Tor client in our experiments
were found to be vulnerable to AS-level correlation attacks. However, due to
aggressive circuit re-use by the Tor client, 58\% of the websites loaded in our
experiments were vulnerable to de-anonymization. When considering Tor clients
used for a mix of applications (Web, BitTorrent, IRC, email, \etc), 30\% of the
circuits were found to be vulnerable.

\item From our historical analysis, we find that the threat faced by Tor clients
has grown. In the context of clients used for loading webpages, we found the
number of vulnerable circuits used by the client increased from 38\% (2010) to
41\% (2015). In the context of clients used for a mix of applications, we found
the number of vulnerable circuits increased more drastically -- from 21\% (2010)
to 35\% (2015). These results show that the threat has been increasing in spite
of a massive growth in the size of the Tor network.
\end{itemize}

\myparab{Evaluating existing defenses (Section~\ref{sec:challenges}).} We
identify five pitfalls in the design of AS-aware Tor clients: (1) a lack of
accurate Internet path data, and not considering (2) the impact of asymmetric
routing on the Internet, (3) the impact of BGP hijack and interception
vulnerabilities, (4) relay-level adversaries, or (5) capacity of Tor relays.
We characterize how these pitfalls impact the security and performance of
existing AS-aware solutions.


 

\myparab{Improving security and performance of AS-aware Tor
(Section~\ref{sec:cipollino}).} Based on our evaluation, we design and construct
\systemname, an AS-aware Tor client which carefully avoids previous pitfalls
while improving security and performance, compared to the current
state-of-the-art. In particular, we show that only 1.4\% of all the webpages
loaded by the \systemname client were vulnerable to AS-level attacks, compared
to 58\% with the vanilla Tor client. Further, \systemname reduces the attack
surface for relay-based adversaries by 80\% relative to the state-of-the-art
AS-aware Tor client (Astoria). Finally, in terms of performance, the \systemname
client achieves median page-load times that are seven seconds faster than the
Astoria Tor client and only 1.6 seconds slower than the vanilla Tor client. 


%% file: background.tex
\section{Background}\label{sec:background}

In this section, we overview the current state of Tor relay selection and
circuit construction algorithms. Then we present our adversary model which
considers active and passive network-level traffic correlation attacks.

\subsection{The Tor anonymity network}\label{subsec:background:tor}

Tor is a low-latency onion routing network that currently consists of 7.1K
relays and has nearly two million daily users \cite{tor-metrics}. When a user
connects to a destination server \via the Tor client, the client typically
establishes the connection using a nested and encrypted three relay
\emph{circuit}. The first relay, called the \emph{entry-relay}, communicates
directly with the Tor user. The last-relay, called the \emph{exit-relay},
communicates directly with the destination server. The key idea is that no
single relay is simultaneously aware of the identities of both, the source 
(Tor user) and destination of the circuit.

\myparab{Tor relay selection.} The three relays in a Tor circuit are selected
according to the following constraints \cite{TorSpec}: (1) no relay may be
selected twice in the same circuit, (2) no two relays belonging to the same
family (advertised by the relays) may be selected as part of the same
circuit, and (3) no two relays belonging to the same /16 subnet may be chosen as
part of the same circuit.

In addition to the above constraints, Tor is (by default) configured to select
an entry-relay from a restricted set of \emph{guard} relays that are stable and
have good performance metrics. When the Tor client is configured to use guards
as entry-relays, it maintains an ordered list of guards and selects the first
usable (online) relay in this list to serve as its entry-relay
\cite{Dingledine-PETS14}. Guards help mitigate the threat of relay-level attacks
such as the predecessor attack \cite{Wright-TISSEC04}, selective
denial-of-service attacks \cite{Borisov-CCS07}, and relay-level correlation
attacks \cite{Wang-Security14}. For the middle- and exit-relay positions in the
circuit, relays are selected based on their available bandwidths. While the
middle-relay is selected from the set of all available relays, exit-relays are
chosen from a smaller subset of relays which have an exit flag. Since relays are
chosen with probability proportional to their available bandwidths, the problem
of overloading small sets of relays is avoided.
 
\myparab{Circuit construction and usage.} Since relays willing to serve as Tor
exits have the ability to specify which ports and IP addresses they are not 
willing to establish connections to, not all circuits constructed by a Tor
client are usable for incoming connection requests. To deal with this, the Tor
client pre-emptively constructs circuits so that at least two are available 
for every destination port seen in the past hour. This allows connection
requests to be served by existing circuits as soon as they are received. In the
event that the client receives a request that cannot be satisfied by any
available circuit, it constructs a new circuit using an exit-relay that can
serve the IP and port specific to that request.

\subsection{Adversary model}\label{subsec:background:model}

\begin{figure}[t]
\centering

\includegraphics[width=0.45\textwidth]{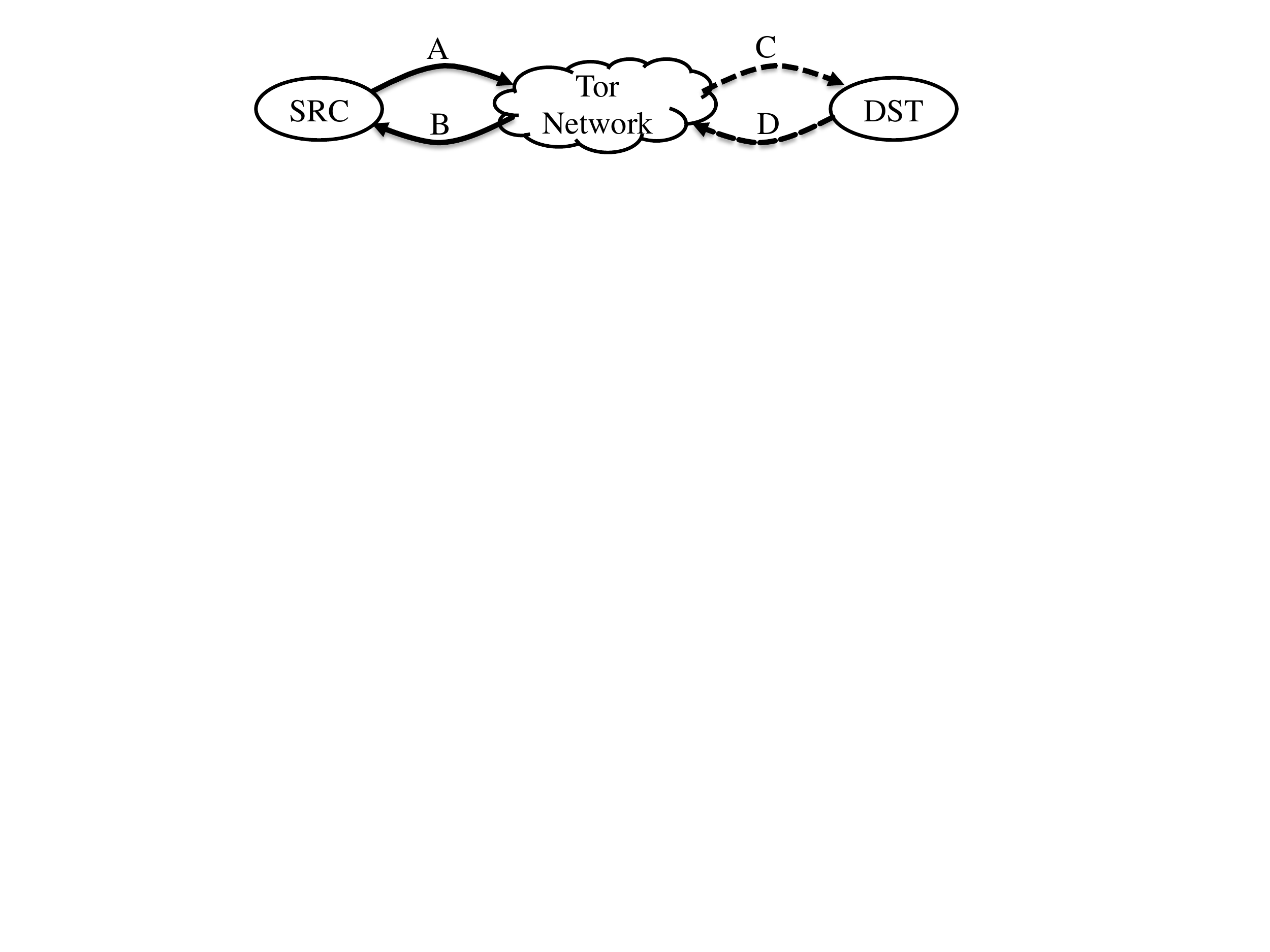}

\caption{Condition required for launching traffic-correlation attacks: An AS
needs to be present on one of the two solid path segments -- \ie path segment A or B
-- and on one of the two dashed path segments -- \ie path segment C or D.}

\label{fig:adversary}

\end{figure}

As a pre-condition to launch network-level correlation attacks, an attacker
(\eg an Autonomous System (AS)) needs to be present on one of the paths
entering the Tor network and one of the paths exiting it. Figure
\ref{fig:adversary} illustrates this condition. Here, to de-anonymize a Tor
client, an AS needs to be present on one of the solid path segments and on one
of the dashed path segments.

More formally, if $P_{SRC \leftrightarrow EN}$ is the set of ASes on the
forward and reverse paths between the Tor client (source) and the selected Tor
entry-relay and similarly, $P_{EX \leftrightarrow DST}$ is the set of ASes on 
the paths between the selected Tor exit-relay and the destination, then we say 
that a Tor circuit is vulnerable to de-anonymization \via traffic-correlation 
if there is some AS $A$ such that:

\begin{equation}\label{eq:adversary} 
A \in \{P_{SRC \leftrightarrow EN} \cap P_{EX \leftrightarrow DST}\} 
\end{equation}

An adversarial AS may satisfy Equation \ref{eq:adversary} through passive or 
active means.

\myparab{Passive adversaries.} An AS may find itself in a position to launch a
traffic-correlation attack simply as a result of the AS-level topology and the
relationships (\ie customer-provider or peer-peer) it shares with other ASes. 
In order to defend against attacks from passive adversaries, it is sufficient to
have an accurate snapshot of the ASes that occur in the sets $P_{SRC
\leftrightarrow EN}$ and $P_{EX \leftrightarrow DST}$ for each choice of $EN$
and $EX$. Given this information, a correlation attack can be avoided by simply
selecting an entry- and exit-relay for which there is no AS $A$ which satisfies
Equation \ref{eq:adversary} (if such an entry- and exit-relay combination
exists). 

\myparab{Active adversaries.} Due to the dynamics and insecurities of the BGP
protocol, ASes may also actively seek to place themselves in a position to
launch traffic-correlation attacks. For example, an AS may hijack or intercept
traffic sent to the prefix associated with the client, entry-relay, exit-relay,
or destination server. Such targeted hijacks and interceptions potentially 
allow adversaries to place themselves on any of the four paths illustrated in
Figure \ref{fig:adversary}.
Defending against such adversaries is more challenging due to need for access to
real-time control-plane data to identify AS that are likely to be hijacking or
intercepting traffic. This is in addition to the snapshots of AS-level paths
required for defending against passive adversaries.

%% file: measurement.tex
\section{Measuring the Threat} \label{sec:measurement}

In this section we describe our methodology for measuring the potential threat
from AS-level adversaries. We use a combination of live experiments on the
current Tor network and simulations that capture a variety of user workloads on
snapshots of the Tor network from 2010 to 2016. Table \ref{Tab:Measurement}
summarizes our experimental setup.

\begin{table}[t] \footnotesize
\centering
\begin{tabularx}{.495\textwidth}{|c|l|X|X|}
\hline
\textbf{Period} & \textbf{Setting} & \textbf{Workload} &  \textbf{Streams
Tested}\\\hline
 2016 & Live & Web model & 215K \\\hline
 2016 & Simulated & mixed model & 12M \\\hline
 2010 - 15 & Simulated & Web model & 145M \\\hline
 2010 - 15 & Simulated & mixed model & 1.9B  \\\hline
\end{tabularx}
\caption{Summary of our experiments to study the threat posed by AS-level
adversaries.}
\label{Tab:Measurement}
\end{table}

%
%
%
%
%
%
%

\subsection{Measurement setting}\label{subsec:measurement:methodology}


We use a combination of live experiments using VPN vantage points and
simulations to understand the threat to Tor in practice, at scale, and over
time.

In our experiments, we consider the fact that due to regional differences in
AS-level topologies, Tor clients in different regions face varying levels
of vulnerability. Therefore, we consider Tor client located in ten countries:
Brazil (BR), China (CN), Germany (DE), Spain (ES), France (FR), England (GB),
Italy (IT), Russia (RU), Ukraine (UA), and the United States (US). This list of
locations was obtained by performing an intersection of the countries with the
largest number of Tor users \cite{tor-metrics} and the countries ranking the
lowest on the Freedom House Internet freedom rankings \cite{freedom-house}.

\myparab{Live experiments with VPNs.} In each live experiment, Crawler
Incantatus \cite{crawler-incantatus} (a Selenium based web-crawler) and the Tor
client were used to load webpages from within each country, using a commercial
VPN.

\myparab{Simulating Tor behavior with TorPS.}  Since the VPN vantage points only
provide us a limited view of each country, in our simulations we considered Tor
clients located in 100 of the most popular (in terms of end-users \cite{aspop})
ASes in each country. 

We use the Tor Path Simulator (TorPS) to analyze the vulnerability of the Tor
network, while considering the massive growth in the Tor network between 2010
and 2015. TorPS is a realistic Python-based Tor simulator which uses archives of
previously published Tor server-descriptors and consensuses from the CollecTor
project \cite{tor-NS} to model historical states of the Tor network. Given (1)
the set of server descriptors corresponding to the period of the experiment and
(2) the set of streams generated by the user (each stream consists of a set of
IP addresses, ports, and connection request times), the TorPS simulator
constructs circuits for each connection request within the stream, according to
a chosen client model (in our case the vanilla Tor client). This allows us to
predict the relays that would have been selected by the Tor client, given a
specific network state from the past.

Each experiment was executed in the live or simulated setting in each of the ten
countries. Additionally, the simulations were used to obtain a picture of the
vulnerability of the Tor client based on network states obtained for the Tor
network between 2010 and 2015. Logs were maintained to track the circuits
established by the Live or Simulated Tor client. 

\subsection{Workload models}
When the Tor client selects relays for a circuit, it may only select exit relays
that have agreed to transport the type of traffic to be sent over the circuit
(\eg some exits restrict commonly abused ports such as port 25 -- SMTP). As a
result, the vulnerability of the Tor client can depend on the applications used
by the Tor user. We consider two different client workloads described below.

\myparab{Web model.} For each experiment using the web user model, 200 websites
were loaded by the Tor client. The list of 200 websites were dependent on the
client location -- \ie comprised of the local Alexa Top 100 sites
\cite{alexa-top} and 100 country-specific sensitive (likely to be blocked or
monitored) webpages obtained from the Citizen Lab testing list repository. In
the case of simulated Tor clients, streams that were used as input to the TorPS
simulator were constructed using the IPs and ports observed in the live
experiments.

\myparab{Mixed (application) model.} For each experiment considering a mixed user
model, we considered clients that used Tor for a mix of Web, P2P (BitTorrent),
e-mail and IRC chat for an hour long period. The  purpose of these experiments
was to understand if the security of the Tor client was affected when users
required connections to non-HTTP(S) ports.

\subsection{Identifying vulnerable circuits} 
To measure the threat posed by network-level attackers, we need to be able to
identify the different networks (\ie ASes) traversed by packets sent between
the Tor entry- / exit-relay and client / destination server. However, ISPs
generally treat their routing information and relationships as trade secrets,
making predictions based on simulations inaccurate. To mitigate this problem, we
use a novel path prediction toolkit -- PathCache~\cite{PathCacheWebsite}. The
main idea behind PathCache is to perform AS-level path prediction  by utilizing
existing publicly available measurements obtained from data-plane measurement
platforms such as RIPE Atlas \cite{ripe-atlas}, iPlane \cite{Madhyastha-OSDI06},
CAIDA Ark \cite{CAIDA-Ark}, and control-plane measurement platforms such as
RouteViews \cite{RouteViews}, RIPE RIS \cite{RIPE-RIS}, and many others. 


In the remainder of our experiments, we consider a circuit constructed by a 
Tor client to be vulnerable if the set of ASes $A$ in Equation
\ref{eq:adversary} is non-empty. Here, we use the PathCache framework to
identify the ASes on $P_{SRC \leftrightarrow EN}$ and $P_{EX \leftrightarrow
DST}$.

\subsection{How vulnerable is the Tor client to AS-level adversaries?}
\label{subsec:measurement:results}

\begin{figure}[t]
\includegraphics[trim=0cm 0.1cm 0cm 2.25cm, clip=true,width=.495\textwidth]
{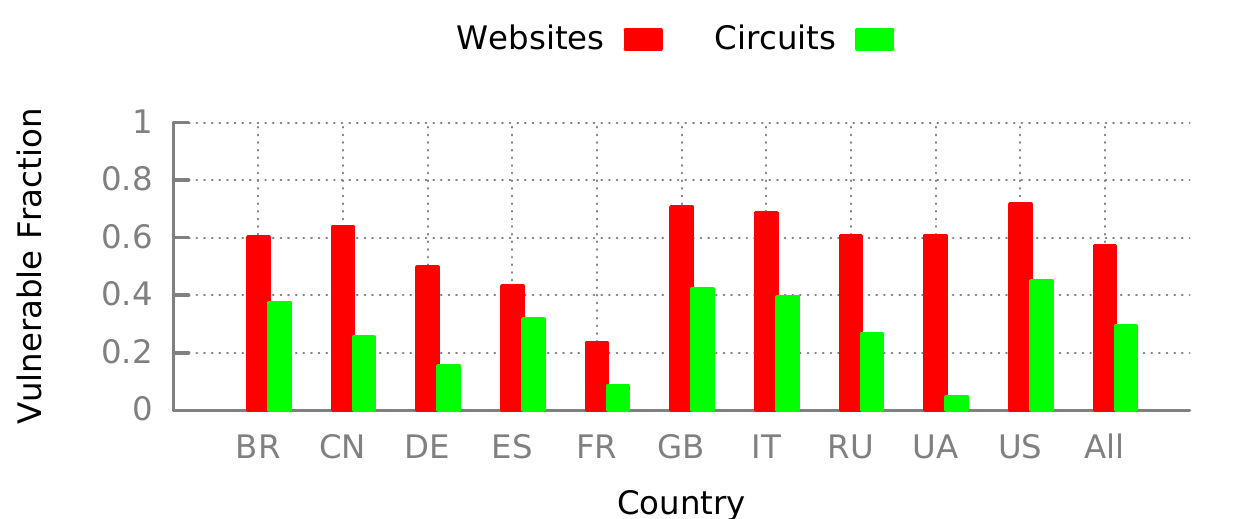}
\caption{Per-country breakdown of fraction of vulnerable websites and circuits
[\textbf{M1}].}
\label{fig:m1-results}
\end{figure}

\myparab{M1: Measuring vanilla Tor's vulnerability to AS-level adversaries (web 
model).} In this experiment we measured the fraction of vulnerable circuits
constructed by the vanilla Tor client and the fraction of websites that use one
of these vulnerable circuits. The results, for each of the ten countries, are 
illustrated in Figure \ref{fig:m1-results}. The experiments were conducted using
a VPN vantage point in each of the ten countries, while loading 200 webpages
(from each). 

\textit{Observation:} 
While only 31\% of the circuits constructed by the Tor client are vulnerable to
AS-level adversaries, we find that due to aggressive circuit re-use and
concentration of websites in a few ASes, that a larger fraction (58\%) of all
websites loaded by the clients end up using a vulnerable circuit.

\begin{figure}[t]
\includegraphics[trim=0cm 0.1cm 0cm 2.25cm, clip=true,width=.495\textwidth]
{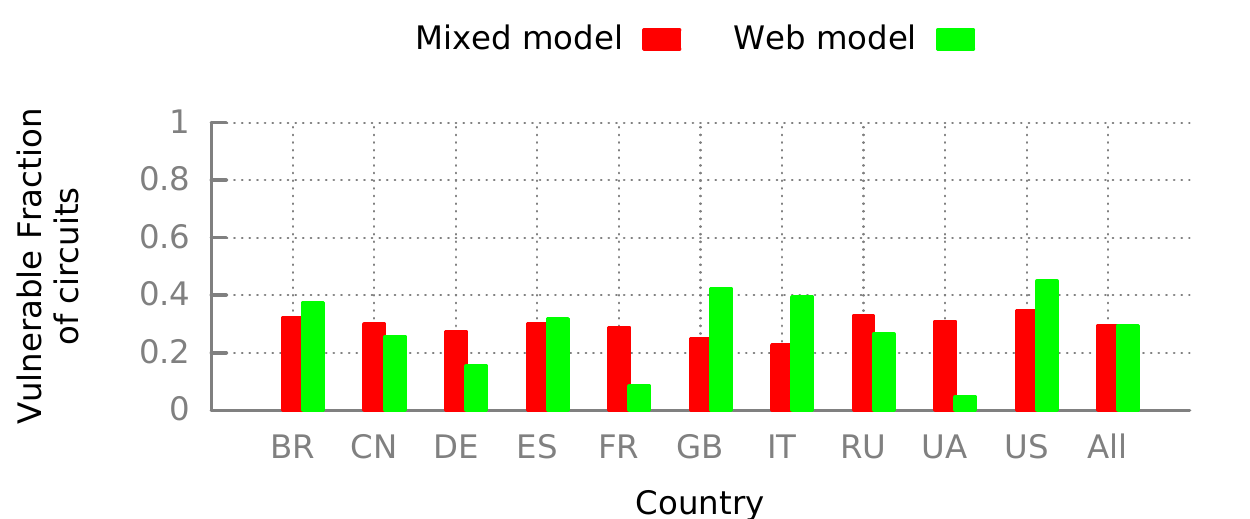}
\caption{Per-country break down of fraction of vulnerable circuits found to be
vulnerable with the Web and mixed user models. [\textbf{M2}].}
\label{fig:m2-results}
\end{figure}

\myparab{M2: Measuring current vulnerability to AS-level adversaries (mixed
model).} In this experiment we measured the fraction of vulnerable circuits
constructed by the vanilla Tor client when it was used for a mix of loading
webpages, sending email, communicating \via IRC chat, and downloading files
using BitTorrent. The results for each of the ten countries are illustrated in
Figure \ref{fig:m2-results}. The experiments were simulated using the TorPS
simulator and a user model based on streams generated by the above applications.
100 of the most populous (in terms of end-users) ASes \cite{aspop} in each of 
the ten countries were selected as Tor client locations.

\textit{Observation:} We find that although the average vulnerability of mixed
application clients (30\%) in the countries is similar to web-only clients
(31\%), the average vulnerability of clients in DE, FR, and UA are most affected
by considering mixed application traffic. This implies that the few exit-relays
that allow communication over non-HTTP(S) ports enable at-least one AS to
perform a traffic correlation attack, given clients located in these countries. 

\subsection{How has the vulnerability evolved as the Tor network has grown?}
\begin{figure}[t]
\includegraphics[trim=0cm 0.1cm 0cm 2.25cm, clip=true,width=.495\textwidth]
{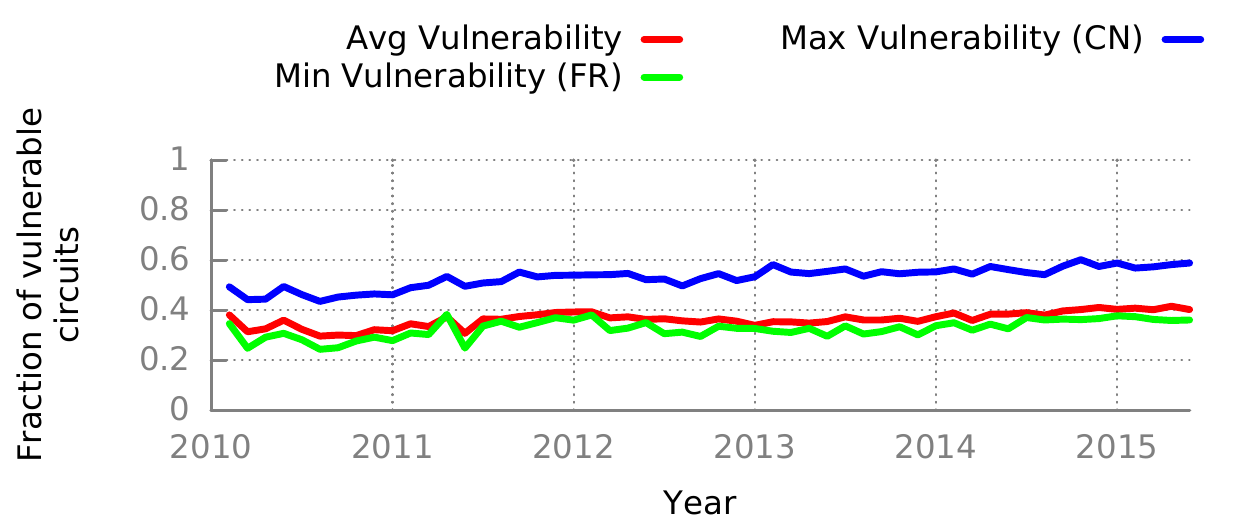}
\caption{The average, current minimum, and current maximum fraction of
vulnerable circuits constructed by vanilla Tor when considering web model
clients located in each of ten countries and the Tor network between 2010 and
2015 [\textbf{M3}].}
\label{fig:m3-results}
\end{figure}

\myparab{M3: Measuring historical vulnerability to AS-level adversaries (web
model).} In this experiment we measured the fraction of vulnerable circuits
constructed by the vanilla Tor client when loading 200 webpages from each of
our ten countries, while considering the changing landscape of the Tor ecosystem
between 2010 and 2015. In each country we consider clients located in the 100
most populous ASes \cite{aspop}. Figure \ref{fig:m3-results} illustrates 
our results. Here, we show the average fraction of vulnerable circuits for 
clients in all 1000 ASes, the country whose 100 ASes had the least average
vulnerability (FR), and the country whose 100 ASes had the highest average
vulnerability (CN).

\textit{Observation:} Most countries have an average of 25-45\% of their
circuits remaining vulnerable to AS-level attackers. China is an exception with
an average of 50-60\% of their circuits remaining vulnerable. Further, in spite 
of the addition of nearly 6K new relays in the Tor network (since 2010), the 
average threat from AS-level adversaries has grown -- from 38\% of all circuits 
being vulnerable in 2010 to 41\% in 2015.

\begin{figure}[t]
\includegraphics[trim=0cm 0.1cm 0cm 2.25cm, clip=true,width=.495\textwidth]
{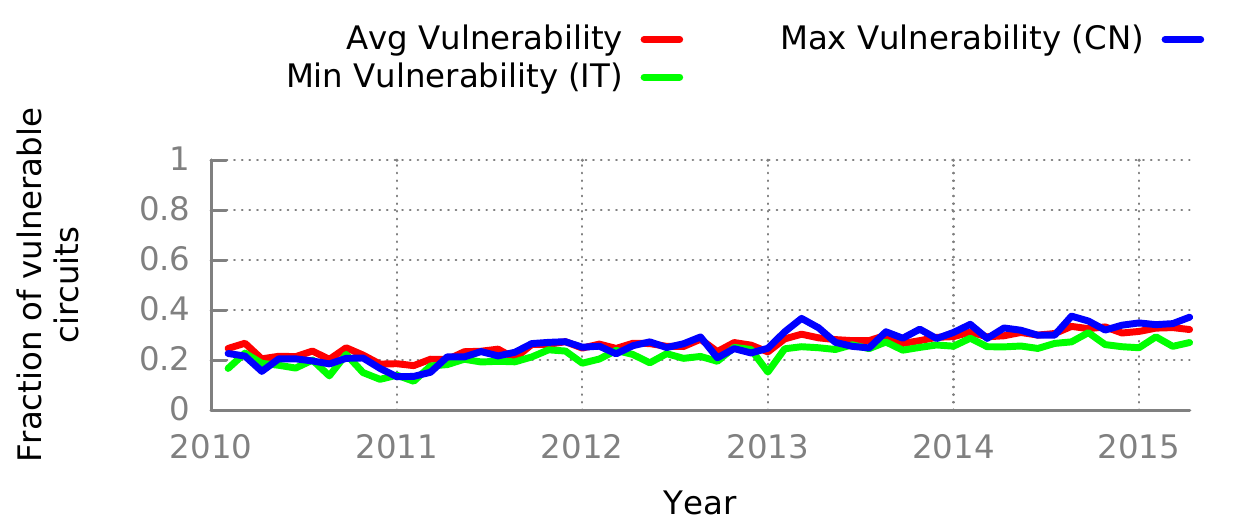}
\caption{The average, current minimum, and current maximum fraction of
vulnerable circuits constructed by vanilla Tor when considering mixed
application model clients located in each of ten countries and the Tor network
between 2010 and 2015 [\textbf{M4}].}
\label{fig:m4-results}
\end{figure}

\myparab{M4: Measuring historical vulnerability to AS-level adversaries (mixed
model).} Here, we use the same settings as experiment \textbf{M3}, only
changing the user model -- \ie while \textbf{M3} calculated the fraction of
vulnerable circuits for users loading 200 webpages in each country, here we
consider users who perform a variety of non-http(s) related communication via
Tor -- \eg IRC, email, BitTorrent, \etc. The results are illustrated in Figure
\ref{fig:m4-results}. 

\textit{Observation:} We find that the threat faced by clients that use Tor for
a mix of non-Web applications is currently slightly lower than web-only Tor
clients, in general. However, the threat has been growing at a significantly
faster rate. We see in the last five years that the average threat (in terms of
vulnerable circuits constructed in the course of our experiments) has increased
from 21\% to 35\%.

\myparab{Discussion.} Our results indicate that the threat from de-anonymization
by AS-level adversaries is significant, regardless of client location and what
the Tor client is used for (web or mixed models). Although the threat faced by
clients used for non-Web purposes is slightly lower, we find that it is growing
at a faster rate than Web-only clients. This is due to the small number of new
non-Web supporting exit-relays being added to the Tor network.

\begin{figure}[t]
\includegraphics[trim=0cm 0.1cm 0cm 2.25cm, clip=true,width=.495\textwidth]
{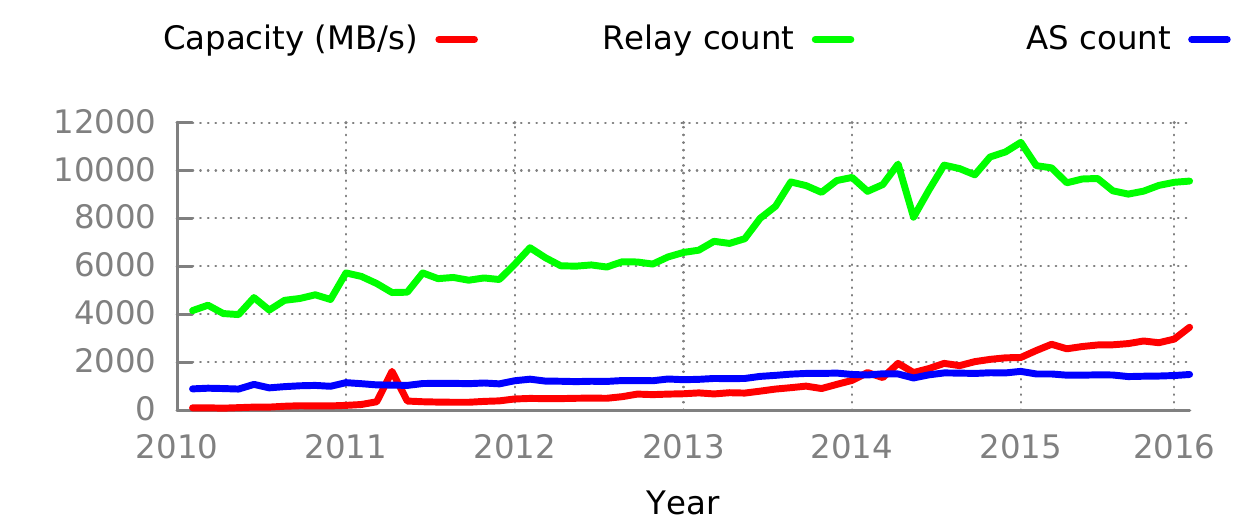}
\caption{The growth of the Tor network in terms of capacity, number of relays,
and number of ASes.}
\label{fig:tor-growth}
\end{figure}

Investigating further into the reason for the growth of the threat from
AS-level adversaries in spite of the massive growth of the Tor network, we find
that while the network has grown, the diversity of the ASes in the network has
not increased. This is illustrated in Figure \ref{fig:tor-growth}. Here, we see
that while the number of relays in the network has grown to nearly 250\% and the 
capacity of the network has grown to over 3000\% of their 2010 values, the
number of ASes in the network has lagged behind (growing to only 160\% of its
2010 value).

\textit{Take-away:} The Tor network faces a fundamental problem when dealing
with AS-level attackers: {the lack of AS-level diversity in the network}. 
In the absence of a specific client-based solution for constructing AS-aware 
circuits, the threat from AS-level attackers is only expected to increase.

%% file: challenges.tex
\section{Pitfalls of AS-aware Tor Clients} \label{sec:challenges}

In this section we survey previous work to identify five common pitfalls
(\textbf{P1} - \textbf{P5}) in the design and construction of AS-aware Tor 
clients. We empirically demonstrate the negative consequences of each.

\subsection{Inaccurate path predictions (P1)} 

The core component of any AS-aware Tor client is its path-prediction toolkit.
The Tor client must accurately identify ASes on the paths from and to the
selected entry- and exit-relays to build circuits that avoid network-level
correlation attacks. Designers of AS-aware clients have three main options for
predicting paths between pairs of ASes:

\myparab{{Data-plane measurements:}} Data-plane measurement tools such as
\texttt{traceroute} allow measurement of exact paths between a source and
destination host. However, this requires control of the source host, which may
not always be possible (\eg it is not possible to traceroute between the
exit-relay and destination server) and has a high latency cost, making it
infeasible for clients to perform on-demand.

\myparab{{Control-plane measurements:}} Paths may also be obtained 
\via control-plane measurement infrastructure such as BGP monitors (\eg
RIPE~\cite{RIPE-RIS}, Routeviews~\cite{RouteViews}). However, they (like
data-plane infrastructure) are limited by the location and peers of the BGP
monitors.

\myparab{{Algorithmic simulations:}} This approach relies on several
simplified assumptions about Internet routing. Typically, algorithmic
simulators use empirically derived AS-level topologies, inferred inter-AS
relationships (\eg customer-provider or peer-peer), and a simplified model of
Internet routing policies (\eg \cite{GaoRexford-TON01, Gill-CCR12}). While
algorithmic simulators are able to predict AS-level paths between any pair of
ASes, their accuracy compares unfavorably with paths obtained from data- and
control-plane measurements. This is due to the incompleteness of AS-level
topologies and the absence of ground-truth while inferring AS relationships.

Table \ref{tab:p1p2} illustrates the design choices of previous efforts to
measure and defend against threats from AS-level attackers. Here, we see 
that all previous work, with the exception of LASTor \cite{Akhoondi-SP12} and 
Juen \etal \cite{Juen-PETS15} relied solely on algorithmic path simulators to 
identify threatening ASes. 

\begin{table*}[t] \footnotesize
\centering
\small
\begin{tabularx}{.985\textwidth}{|X|p{.5in}|p{.5in}|p{.65in}|p{.7in}|X|X|X|}
\hline
\multirow{1}{1cm}{\textbf{ }} & \multicolumn{3}{c|}{\textbf{Path Prediction
Approach (P1)}} & {\textbf{Asymmetric Routes}} & {\textbf{BGP
Insecurities}} & \multirow{2}{.7in}{\textbf{Relay-level attacks}} &
\multirow{2}{.6in}{\textbf{Load balancing}}
\\\cline{2-4} &

\textbf{Data-plane} & \textbf{Control-plane} & \textbf{Simulations} &
\textbf{(P2)}& \textbf{(P3)} & \textbf{(P4)} & \textbf{(P5)}
\\\hline

Feamster \& Dingledine \cite{Feamster-WPES04} & X & X &
$\surd$ \cite{Mao-PER05,Gao-TON01} & $\surd$ & X & X & X
\\\hline

Edman \& Syverson \cite{Edman-CCS09} & X & X & $\surd$ 
\cite{Qiu-GLOBECOM06, Gao-TON01} & $\surd$ & X & X & X
\\\hline

LASTor \footnotemark \cite{Akhoondi-SP12} & X & $\surd$ & X & X & X & X & X
\\\hline

Johnson \etal \cite{Johnson-CCS13} & X & X & $\surd$ 
\cite{Qiu-GLOBECOM06, Gao-TON01} & X & X & $\surd$ & X
\\\hline

Juen \etal \cite{Juen-PETS15} & $\surd$ & X & $\surd$ 
\cite{Qiu-GLOBECOM06, Gao-TON01} & X & X & X & X
\\\hline

Astoria \cite{Nithyanand-NDSS16} & X & X & $\surd$ 
\cite{GaoRexford-TON01, Gill-CCR12} & $\surd$ & X & X & $\surd$
\\\hline

\systemname [This paper] & $\surd$ & $\surd$ & $\surd$ 
\cite{GaoRexford-TON01, Gill-CCR12} & $\surd$ & $\surd$ & $\surd$ & $\surd$
\\\hline

\end{tabularx}

\caption{Comparison of the measurement methodologies and defense contributions
of the state-of-the-art. X indicates the corresponding criteria was not
considered and $\surd$ indicates that it was. \textbf{(P1-P5)}}
\label{tab:p1p2}

\end{table*}
\footnotetext{The client was not made available by the authors after multiple
requests. We try to objectively evaluate the paper based on descriptions in the
text.}

\begin{figure}[t]
\includegraphics[trim=0cm 0.1cm 0cm 2.25cm, clip=true,width=.495\textwidth]
{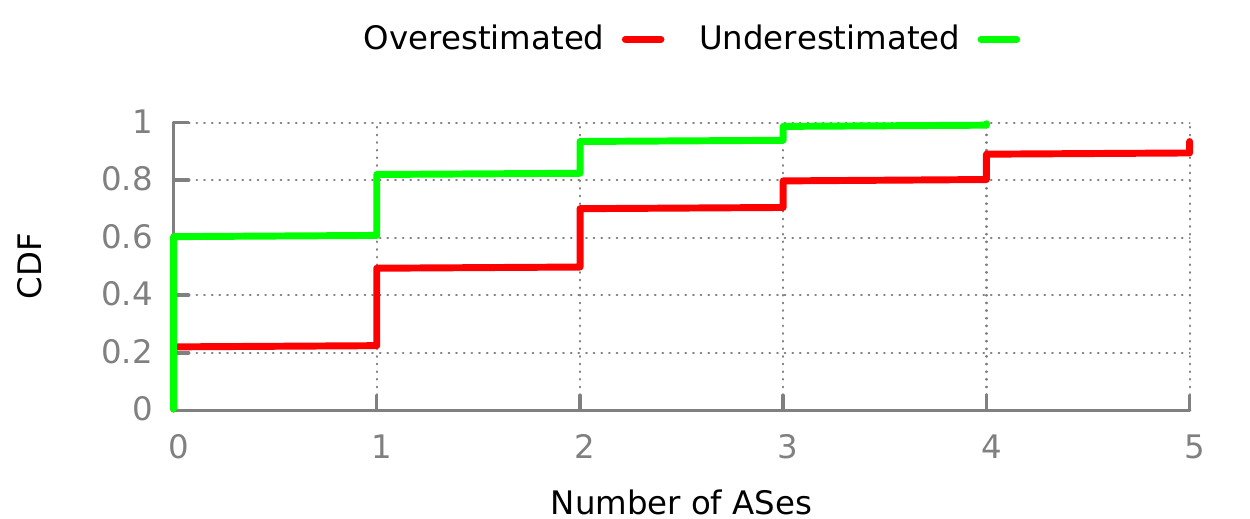}
\caption{Number of ASes over or under-estimated by the state-of-the-art
algorithmic simulator [\textbf{P1}].}
\label{fig:bgpsim-accuracy}
\end{figure}

To understand the impact of inaccurate path predictions, we test the accuracy of
the state-of-the-art simulator \cite{Gill-CCR12} which relies on the Gao-Rexford
routing model \cite{GaoRexford-TON01}. For our experiment, 225 pairs of
exit-relay and destination ASes were chosen from the circuits constructed by the
vanilla Tor client in our VPN experiment (\textbf{M1}). For each pair, a
traceroute was executed from the  AS containing the exit-relay (vantage points
were obtained using RIPE Atlas probes). IPs from each traceroute hop were
resolved to their ASes using up-to-date BGP announcement data to produce
AS-level paths. These AS-level paths were compared with the AS-level paths
obtained by the algorithmic simulator. Figure~\ref{fig:bgpsim-accuracy} shows
the result of comparing measured with simulated paths. We find that
straightforward application of simulation can lead to over estimating the number
of ASes present 80\% of the time. Worse yet, 40\% of the time simulated paths
actually miss ASes contained in the paths, potentially leaving the client
vulnerable to traffic-correlation attacks.


\subsection{Ignoring route asymmetry (P2)}

Recent work by Sun \etal \cite{Sun-Security15} demonstrated, \via high accuracy
AS-level correlation attacks on the Tor network, that the threat from AS-level
attackers was higher than previously anticipated. This is primarily because of
two factors: Adversaries can (1) exploit the asymmetry of routing on the
Internet -- \ie exploit the fact that their presence on the forward- or
reverse-paths at either end of the network is sufficient to launch an attack and
(2) perform manipulation of routes \via BGP hijacks and interceptions to place
themselves on targeted paths. In this section, we consider the impact of
adversaries on asymmetric routes. In the following section, we discuss the
impact of  BGP hijacks on Tor.

From Table \ref{tab:p1p2} we find that the possibility of asymmetric routes was
considered in several previous works. However,  defending against these
attackers is challenging since it requires knowledge of reverse network paths,
many of which cannot be measured directly. This compounds {\bf P1} for AS-aware
Tor clients as these paths need to be predicted as well.


\begin{figure}[t]
\includegraphics[trim=0cm 0.1cm 0cm 2.25cm, clip=true,width=.495\textwidth]
{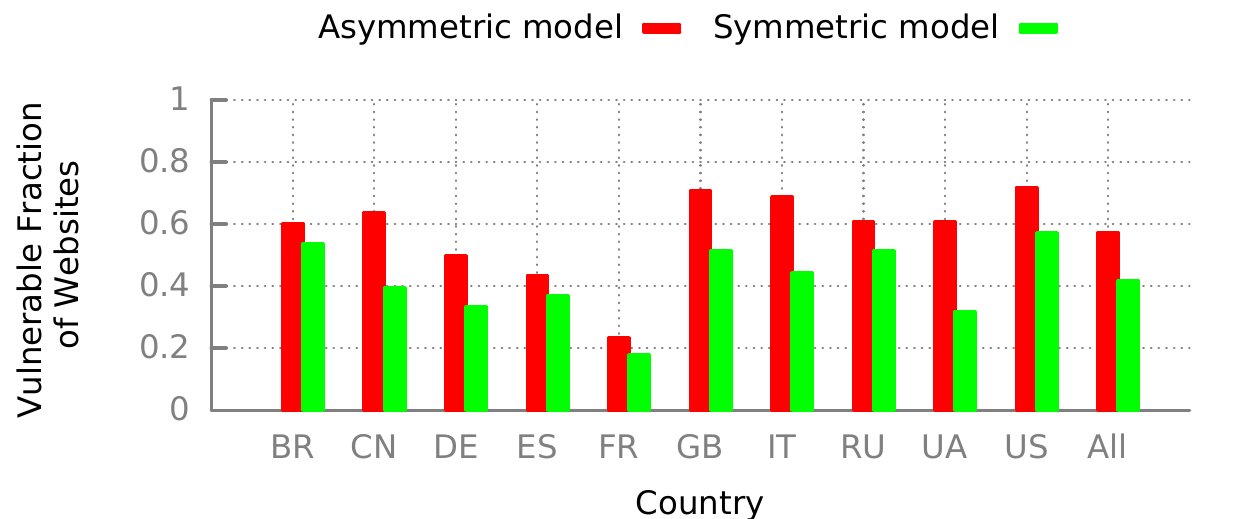}
\caption{Fraction of websites using vulnerable circuits against a symmetric and
asymmetric adversary [\textbf{P2}].}
\label{fig:routing-p2}
\end{figure}

We measure the consequences of not considering an adversary that exploits
asymmetric paths. To do so, we repeat experiment \textbf{M1}, but this time we
only consider an attacker that can exploit only forward paths -- \ie we say that
a circuit is vulnerable to de-anonymization if there is some AS $A$ such that: 
$A \in$ $\{P_{SRC \rightarrow EN} \cap P_{EX \rightarrow DST}\}$.
Figure \ref{fig:routing-p2} compares the fraction of websites marked as
vulnerable against a forward-path exploiting (symmetric) adversary model with
our (asymmetric) adversary model. We find that operating under the assumption of
symmetric routing (\ie considering only forward-path exploiting adversaries)
results in significant threat under-estimation, with circuits to 17\% of all
websites identified as safe when they were in fact vulnerable.

%

\subsection{Ignoring active BGP attacks (P3)}
The potential for BGP hijacks and interceptions to compromise Tor traffic was
highlighted by Sun~\etal \cite{Sun-Security15}. In this section, we measure how
vulnerable Tor relays are to BGP hijacks and interceptions by sets of malicious
ASes. For this experiment, we considered 10K pairs of (source, entry) ASes and
10K pairs of (exit, destination) ASes. The source ASes were randomly selected
from the 1000 popular ASes (100 in each of ten countries) used in experiments
\textbf{M2-M4} while the entry and exit ASes were selected from the set of all
Tor entry and exit relays, respectively. Destination ASes were randomly chosen
from the set of all destination ASes seen in experiment \textbf{M1} (when
loading 200 webpages in each of ten countries). For our adversary (\ie ASes
attempting to launch hijack and interception attacks), we selected the 16
malicious ASes identified in previous work \cite{Konte-SIGCOMM15} as popular
ASes for hosting illegal content, botnet C\&C servers, and other malicious
resources. 

For each pair of ASes we use heuristics from Goldberg \etal
\cite{Goldberg-SIGCOMM10} to check which of the 16 malicious ASes is capable of
hijacking or intercepting traffic between the pair of ASes. 

\begin{figure}[t!]
\centering
\begin{subfigure}{.5\textwidth}
\includegraphics[trim=0cm 0.1cm 0cm 2.25cm, clip=true,width=\textwidth]
{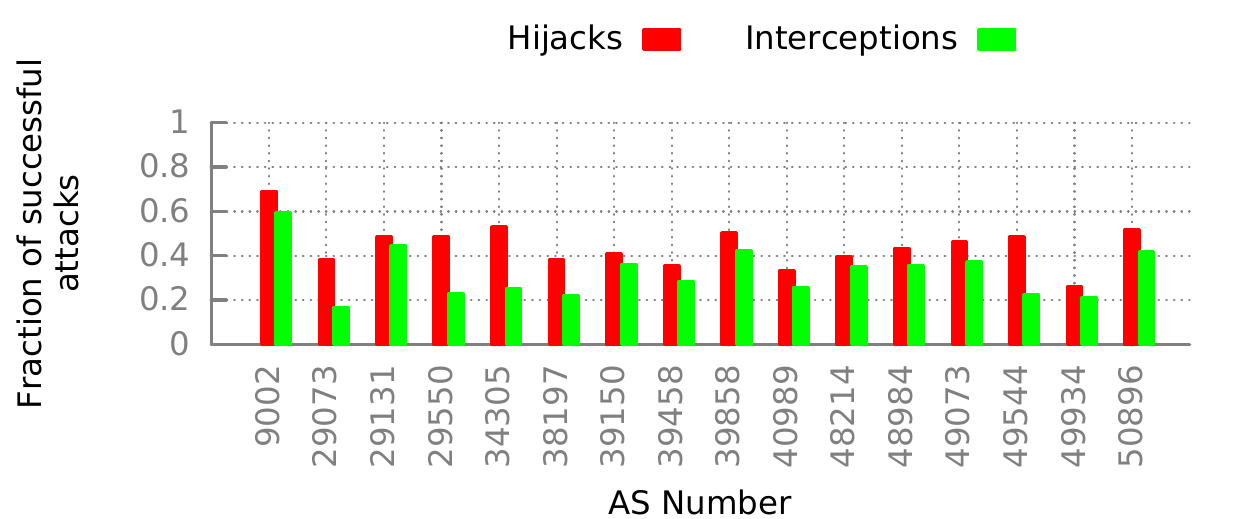}
\caption{Fraction of successful hijack and interception attacks by each chosen
malicious AS against selected Tor relays.}
\label{fig:as-attacks-p2}
\end{subfigure}
\begin{subfigure}{.5\textwidth}
\includegraphics[trim=0cm 0.1cm 0cm 2.25cm, clip=true,width=\textwidth]
{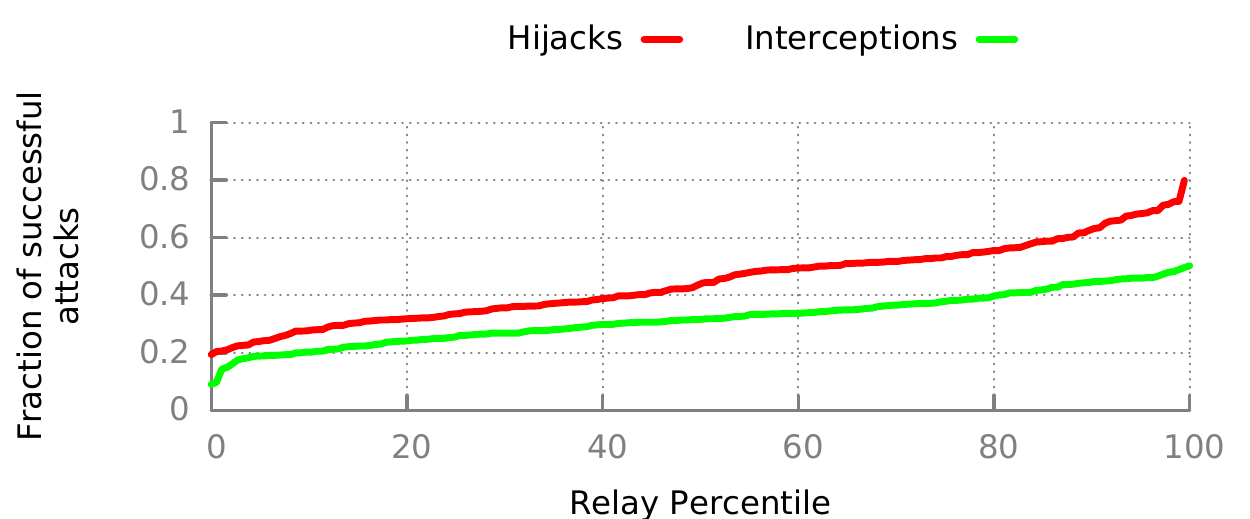}
\caption{Fraction of successful hijack and interception attacks by 16 malicious
ASes against each chosen Tor relay.}
\label{fig:relay-attacks-p2}
\end{subfigure}

\caption{Threats from omitting BGP insecurities in the adversary model
[\textbf{P3}].}
\end{figure}

We first characterize the ability of the malicious ASes to hijack traffic for a
chosen path. Figure \ref{fig:relay-attacks-p2} demonstrates the hijack and
interception success rates of each of the 16 ASes considered in this experiment.
Here we see that two ASes -- ASN 9002: RETN (UA), ASN 29131: RapidSwitch (GB) --
achieve high hijack and interception success rate of nearly 50\%. The case of
ASN 9002 can be explained by its high customer cone size (3271 customer ASes).
On the other hand ASN 29131 is a smaller AS with only one customer AS, however,
it peers with seven other large ASes having an AS rank under 1K (based on
customer cone sizes). 

Next we wanted to understand how vulnerable given Tor relays are to attack.
Specifically, Figure \ref{fig:as-attacks-p2} shows the fraction of
hijack/interception attempts were successful for the relays in ascending order.
Each one of the relays we consider is susceptible to at least 20\% of hijacks
and 12\% of interception attempts.


\subsection{Increasing risk of relay adversaries (P4)}

We argue that a client which utilizes a smaller number of relays to serve
connection requests, over a period of time, is less likely to encounter a
malicious relay in the Tor network. Thus, a Tor client that uses a smaller
number of relays is more secure against adversarial relays.

We observe that many proposed defenses \cite{Feamster-WPES04, Edman-CCS09,
Akhoondi-SP12, Nithyanand-NDSS16} do not consider the impact of AS-aware relay
selection on the security of the client against relay-level adversaries. This is
problematic because many AS-aware clients build circuits on a per-destination
basis, as opposed to reusing a smaller set of existing circuits. This results
in them leveraging a large set of relays over time.


To illustrate the impact of destination-based circuits, we conduct an experiment
using the Astoria Tor client \cite{Nithyanand-NDSS16}. The Astoria Tor client
performs on-demand circuit construction for each new destination AS that it
encounters, while re-using valid (live) circuits for previously seen ASes. In
this experiment, we use our VPN end points and crawler to load the 200 pages of
the Web user model (\ie the same settings as \textbf{M1}) using Astoria and the
vanilla Tor clients. We log the number of unique relays utilized by the Astoria
and Tor clients to serve the page loads. We find that circuits generated by
Astoria utilize nearly five times more relays than the vanilla Tor client -- \ie
Astoria utilized {3,104} unique relays compared to the {623} relays used by the
Tor client. This is a drastic increase in the potential for encountering
malicious relays over the vanilla Tor client. 



\subsection{Overloading relays in the Tor network (P5)} 
We find that much of previous work \cite{Feamster-WPES04, Edman-CCS09,
Akhoondi-SP12} does not consider the capacities of relays chosen as part of
AS-aware circuits. We argue that relay capacity is important to consider to
prevent custom relay selection schemes from overloading low-capacity relays
and reducing performance across the population of Tor users. 

As an example of the impact of ignoring relay capacities, Wacek \etal 
\cite{Wacek-NDSS13} performed a study to analyze the throughput of various Tor
relay selection strategies and found that: (1) strategies that ignored 
relay capacities had significant drops in both, client and network throughput
and (2) while LASTor had better performance than vanilla Tor when considering
round-trip times on established circuits, the throughput of the client when 
used for page loads was 70\% less than the Tor client (compared to 25\% more
than Tor as demonstrated in original work by Akhoondi \etal).

The reason for this large disparity in performance reported in the two
evaluations is due to Akhoondi \etal \cite{Akhoondi-SP12} only sending HTTP HEAD
requests in their experiments (as opposed to downloading complete webpages or
documents). In addition to being unrepresentative of typical web traffic, such
evaluations do not sufficiently stress all the relays chosen as part of a
circuit and as a result do not reveal the issues associated with
capacity-agnostic relay selection.

\Pnote{I thought we had a result of astoria with and without load balancing that
showed worse performance. fi that made the cut for NDSS can we reference it
here? Or can we resurrect it if it wasn't in the NDSS paper?}

\rish{Nope. It never made it to the paper. Can't find it in the repo or on the
server either. The closest thing we have is the LP manipulation results with the
variation of the $\alpha$ parameter we played around with for a while. Even
that, I cannot find the data, only the plot :(}


%% file: cipollino.tex
\section{The Cipollino Tor Client} \label{sec:cipollino}

Based on the pitfalls we identified in the prior section, we design \systemname,
an AS-aware Tor client that uses state-of-the-art network measurements and
optimizations to mitigate the pitfalls. Table~\ref{tab:cipollino} summarizes
how \systemname addresses each of the pitfalls described in Section
\ref{sec:challenges}. We elaborate on each in the following sections.

\begin{table}[t]
{\small 
\begin{tabular}{|p{0.25\textwidth}|p{0.2\textwidth}|}
\hline
{\bf Pitfall} & {\bf Solution} \\
\hline
\hline
P1. Simulated network paths & PathCache empirical data \\
\hline 
P2. Ignoring route asymmetry & Including reverse paths in decision making\\
\hline
P3. Ignoring BGP hijacks & Realtime BGP data\\
\hline 
P4. Increasing risk of relay adversaries & Reuse safe circuits between destinations\\
\hline
P5. Overloading Tor relays & Load balance across safe circuits\\
\hline
\end{tabular}
}
\centering
\caption{Overview of how \systemname mitigates the pitfalls of prior AS-aware
Tor clients.}\label{tab:cipollino}
\end{table}

\subsection{Improving path prediction (P1)}

We reduce our dependence on algorithmic simulators by using PathCache
\cite{PathCacheWebsite} -- a system that aggregates existing data and control
plane measurements to predict paths. We fall back to simulations only when a
path query cannot be answered using measurement data. Repeated querying of the
PathCache server every time a circuit needs to be built is (1) time consuming
and (2) reveals destinations of interest to a third party (\eg PathCache
server). To avoid this, the \systemname client subscribes to daily updates of
the routing graphs maintained by the PathCache server and  locally computes
paths between ASes. This is beneficial for two other reasons:

\begin{enumerate}

\item Offline verification of paths: Since the meta-data for each edge in the
routing graphs maintained by PathCache includes information regarding the source
of the edge (\ie to indicate the edge was observed in a traceroute from the RIPE
Atlas network, control-plane data from RouteViews, \etc) and the measurement ID
corresponding to the source. This is useful for the client to verify the
authenticity of of a random subset of the daily updated paths supplied by the
\systemname aggregator.

\item Low communication overhead: The routing graph updates are between 5-15
MB/day. This is feasible for clients in most settings. Additionally, it allows
clients to identify safe circuits even if the PathCache server is not
immediately reachable.

\end{enumerate}

To understand the benefits of PathCache, we evaluate PathCache on two criteria:
(1) accuracy of predicted paths and (2) the fraction of paths where PathCache is
able to answer using empirical data (\vs simulations).

\begin{figure}[t]
\includegraphics[trim=0cm 0.1cm 0cm 2.25cm, clip=true,width=.495\textwidth]
{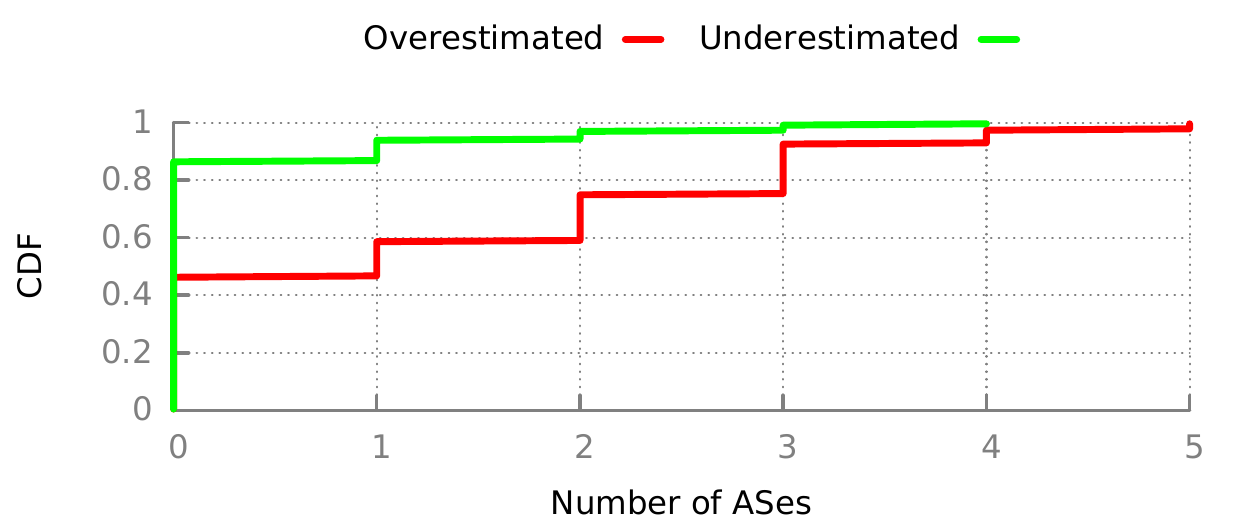}
\caption{Number of ASes over or under-estimated by PathCache, when compared to
exact AS-level paths obtained by traceroutes [\textbf{P1}].}
\label{fig:PC-accuracy}
\end{figure}

We measure the number of ASes over- or under-estimated when compared with 225
traceroutes that were not already aggregated by PathCache. Figure
\ref{fig:PC-accuracy} illustrates the results of this experiment. We find that
PathCache is significantly more accurate than the state-of-the-art algorithmic
simulator (cf. Figure~\ref{fig:bgpsim-accuracy}). Most importantly, with 84\% of
all paths having no missing ASes (no underestimations), PathCache is much less
likely to create vulnerable circuits due to incorrect path predictions.

To understand how often PathCache is able to predict paths using empirical data,
we queried PathCache for paths between (1) 1,000 source ASes (100 of the most
populous ASes in each of the ten countries) and the ASes of all entry-relays in
the Tor network (265K path queries) and (2) between the ASes of all exit-relays
in the Tor network and all the destination ASes seen  in our 2,000 web-page
loads (312K path queries). Table \ref{tab:coverage} shows the percentage of
paths that were predicted by PathCache using empirical data. Here we see that
PathCache is able to achieve reasonable coverage when considering high capacity
entry- and exit-relays (34-36\%). This implies a higher accuracy of paths
predicted for organically generated Tor circuits, as the Tor client will tend to
use these higher capacity relays.

\begin{table}[t] \footnotesize
\centering
\begin{tabularx}{.495\textwidth}{|l|X|X|}
\hline
\textbf{N} & \multicolumn{2}{c|}{\textbf{Coverage (Percentage)}} \\\cline{2-3}
\textbf{} & $SRC \leftrightarrow EN$ & $EX \leftrightarrow DST$ \\\hline
10 & 36.6 & 34.0 \\\hline
25 & 32.1 & 32.4 \\\hline
50 & 27.6 & 31.3 \\\hline
100 & 23.2 & 29.4 \\\hline
\end{tabularx}
\caption{Percentage of paths predicted by PathCache when considering only the
top N percentile of relays (by bandwidth) [\textbf{P1}].}\label{tab:coverage}
\end{table}

\begin{figure}[t]
\includegraphics[trim=0cm 0.1cm 0cm 2.25cm, clip=true,width=.495\textwidth]
{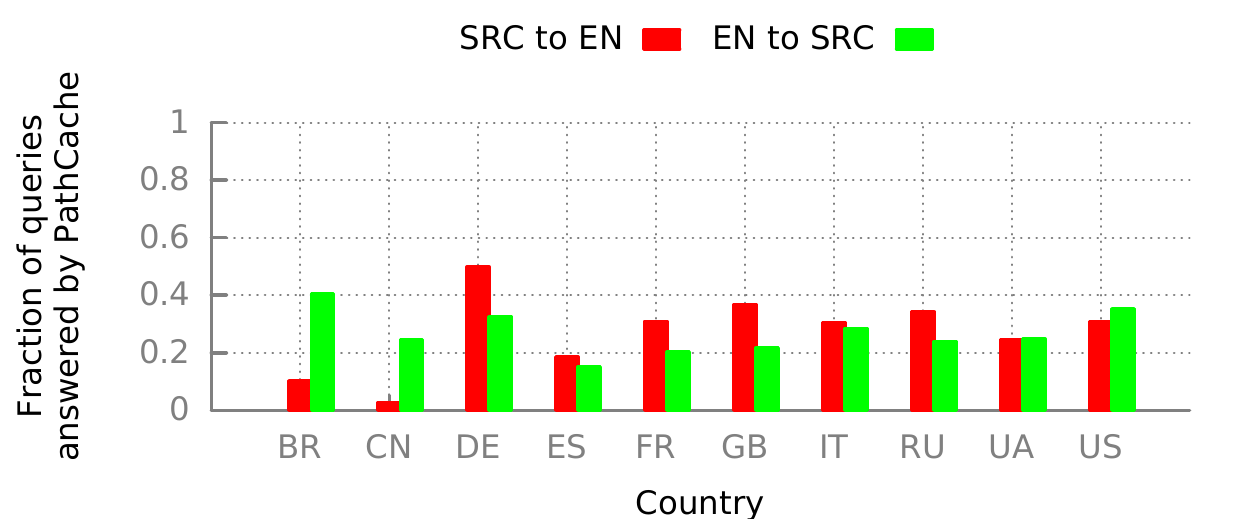}
\caption{Per-country breakdown of $SRC\leftarrow EN$ and $EN\leftarrow SRC$
paths predicted by PathCache [\textbf{P1}].}
\label{fig:PC-coverage}
\end{figure}

In Figure \ref{fig:PC-coverage} we see the per-country breakdown of the fraction
of path requests satisfied by PathCache. Interestingly, we see BR and CN in
particular having a very small fraction of paths between their 100 AS sources
and the Tor entry-relays. We speculate that this is due to blocking of
communication with Tor entry-relays in these countries. This prevents
traceroutes (that PathCache uses as a basis for path prediction) from
successfully traversing paths from client ASes in these countries to Tor entry
relay ASes. 

Depending on client location, PathCache is able to answer between 15-50\% of all
queries issued to it by the Tor client. Importantly, the paths returned for
these queries are unlikely to under-estimate the presence of an AS.
Additionally, coverage increases significantly when considering higher capacity
Tor relays. These factors make it a good alternative to relying on simulations
for path prediction. 

\subsection{Considering active adversaries that can exploit asymmetric routes
(P2-P3)}

\systemname considers an adversary model that includes the possibility of ASes
exploiting (1) asymmetric routes and (2) BGP insecurities. To explain how we
deal with such adversaries, we describe how \systemname verifies the safety of a
given circuit below.

\begin{packeditemize}
\item \emph{Mapping destination IP addresses to ASNs and prefixes:} Given a
circuit and a destination IP address, the \systemname client first uses an
up-to-date offline IP to ASN database (based off of BGP announcements) to obtain
the AS numbers associated with the network of the client, entry-relay,
exit-relay, and requested destination IP. This database (sourced and updated by
CAIDA) is supplied and updated by the PathCache daily updates. 

Following this, \systemname generates two pairs of ASes and two pairs of
prefixes -- ($AS_{EN}$, $AS_{SRC}$), ($AS_{EX}$, $AS_{DST}$), ($Pre_{EN}$,
$Pre_{SRC}$), and ($Pre_{EX}$, $Pre_{DST}$).

\item \emph{BGP anomaly detection:} In order to detect hijacks and interceptions
in near-real-time, \systemname receives hourly (customizable in the client
configuration) feeds from BGPStream \cite{CAIDA-BGPStream} of current BGP
routing anomalies. In particular, BGPStream produces a live stream of ongoing
Multiple Origin AS (MOAS) anomalies. MOAS anomalies, which occur when a prefix
is being announced by multiple origin ASes. We use MOAS as an indicator of
potentially anomalous routing behavior as a proof of concept. Beyond the scope
of this paper we are working to develop more accurate detection methods for
hijacks and interceptions which could be incorporated into
\systemname \cite{caida-hijacks}. This feed of ASes is used to identify ASes
that are likely to be hijacking or intercepting traffic to any of the prefixes
in the previously generated pairs -- ($Pre_{EN}$, $Pre_{SRC}$) and ($Pre_{EX}$,
$Pre_{DST}$). 

Any AS $X$ that is suspected to be hijacking or intercepting traffic to the
prefix associated with the entry-relay is added to the set $H_{EN}$. Similarly, the
sets $H_{EX}$, $H_{SRC}$, and $H_{DST}$ are populated. 

\item \emph{Path prediction:} The \systemname client uses the locally stored
PathCache destination based graphs to obtain the set of ASes on the  $SRC
\leftrightarrow EN$ and $EX \leftrightarrow DST$ paths. 
Additionally, the ASes occurring on the paths between $H_{EN} \leftarrow SRC$
and $EN \leftarrow H_{EN}$ are added to $SRC \leftrightarrow EN$. This accounts
for all ASes that are able to view traffic characteristics in the event of a
successful interception (and hijack) of traffic to $EN$. The same process is
repeated for $H_{EX}$, $H_{SRC}$, and $H_{DST}$.

\item \emph{Circuit safety marking: } After the paths are computed, a circuit is
marked as safe iff the sets $SRC \leftrightarrow EN$ and $EX \leftrightarrow
DST$ have no intersection. 
\end{packeditemize}

The circuit safety verification procedure shows that \systemname does not mark a
circuit as safe to serve a given destination unless there are no ASes that are
in a position to view traffic characteristics at either end of a circuit, after
accounting for route asymmetry and potential hijacks. 

\subsection{Pre-building and re-using circuits (P4)} 

To reduce the number of relays it uses, \systemname  employs a circuit
pre-building strategy similar to the vanilla Tor client.  \systemname
pre-emptively constructs a fixed (and configurable) number of circuits. In
addition to the benefit of reduced utilization of relays, two other arguments
for pre-emptive circuit construction come from the following observations drawn
from previous work by Nithyanand \etal \cite{Nithyanand-NDSS16}: 

\begin{packeditemize}
\item For over 50\% of all client locations and destination ASes considered,
at-least 50\% of all possible entry- and exit-relay combinations were safe from
correlation attacks by AS-level adversaries. Therefore, by pre-building a number
of circuits, we are very likely to find at least one safe circuit for a given
destination AS. 

\item Constructing a new circuit is significantly more expensive than verifying
the safety of an existing circuit -- \ie due to the need for estimating the
paths between all possible (source, entry-relay) and (exit-relay, destination)
pairs. Therefore, by pre-emptively constructing circuits, \systemname reduces
the need to construct on-demand destination-aware circuits. 
\end{packeditemize}

\begin{figure}[t]
\includegraphics[trim=0cm 0.1cm 0cm 2.25cm, clip=true,width=.495\textwidth]
{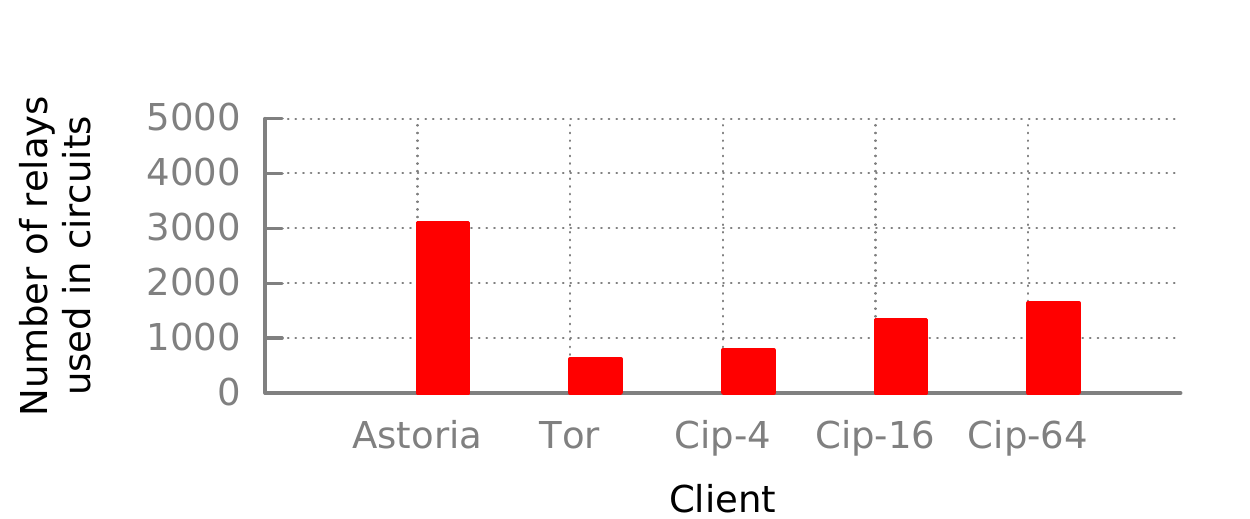}
\caption{Number of unique relays used in circuits constructed by each client
while loading 200 webpages in each of ten countries [\textbf{P4}].}
\label{fig:prebuild}
\end{figure}

To understand how circuit pre-building affects the number of relays used by
\systemname, we consider the 200 Web pages loaded in the Web user model with
\systemname   configured to pre-build and always maintain 4, 16, and 64 live and
usable circuits. Figure \ref{fig:prebuild} compares the number of relays used in
each setting with the vanilla Tor client and Astoria. When  \systemname  is
configured to only pre-build and maintain 4 active circuits, it utilizes 786
relays (compared to the 623 relays used by Tor). This is significantly lower
than Astoria (3104 relays).

\begin{figure}[t]
\includegraphics[trim=0cm 0.1cm 0cm 2.25cm, clip=true,width=.495\textwidth]
{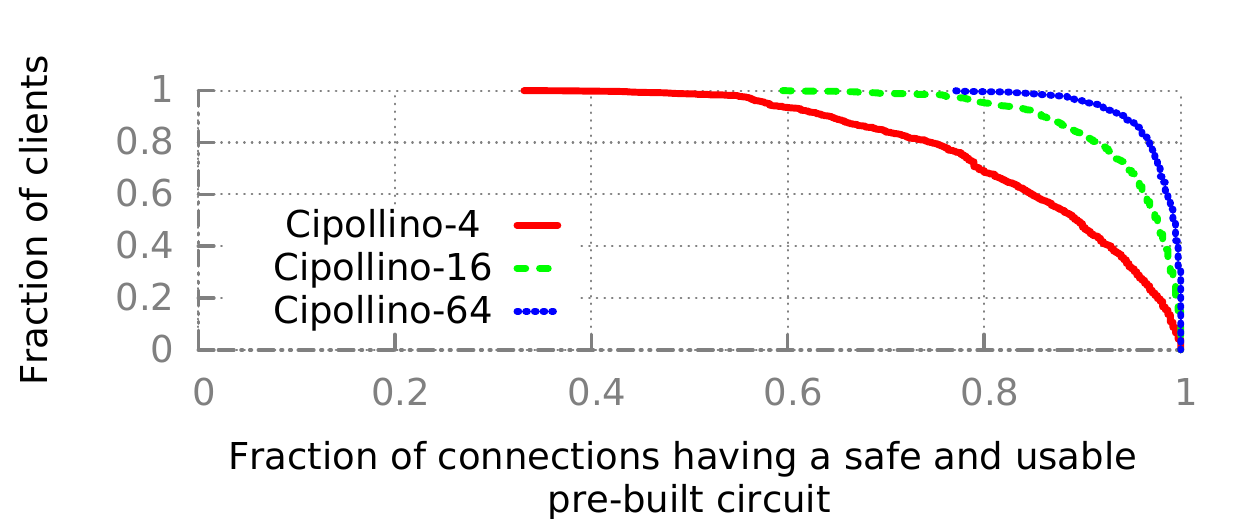}
\caption{Distribution of the fraction of connection requests that were able to
find a safe and usable circuit from 4, 16, and 64 circuits pre-built be the
\systemname client [\textbf{P4}].}
\label{fig:pbc-safety}
\end{figure}

Figure \ref{fig:pbc-safety} also illustrates that pre-building circuits
results in the need for constructing fewer on-demand and destination-aware
circuits. In this experiment, 1,000 \systemname clients were simulated (with
locations in the 100 most populous ASes in each of ten countries) and issued
connection requests for destinations associated with 200 country-specific
webpages. Here we see that 50\% of the clients were able avoid on-demand
circuit construction for at-least 86\% of the connection requests, when just
four circuits were prebuilt. 

\begin{figure}[t]
\includegraphics[trim=0cm 0.1cm 0cm 2.25cm, clip=true,width=.495\textwidth]
{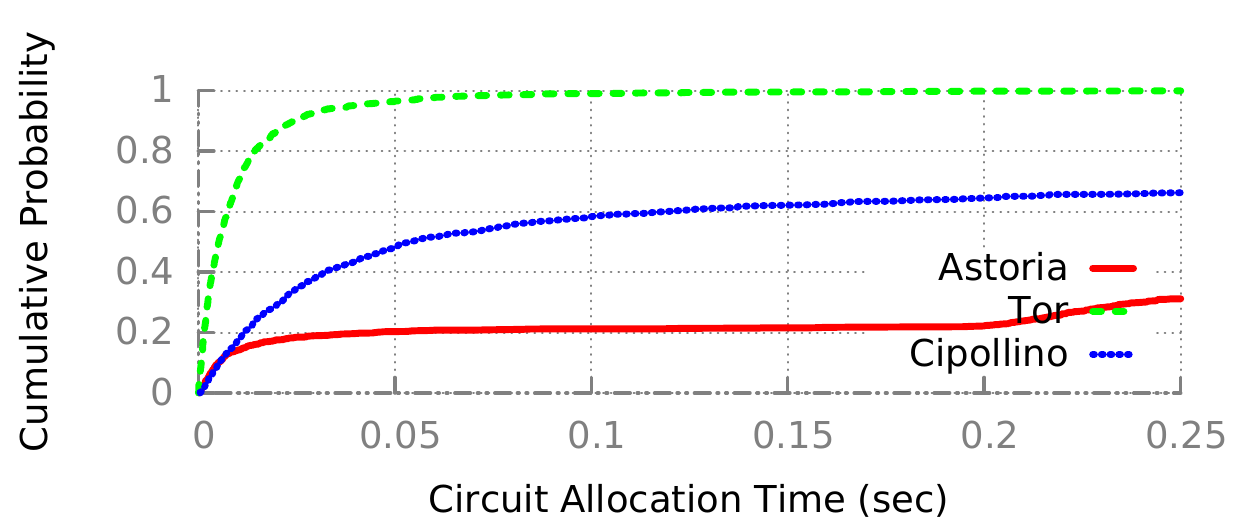}
\caption{Distribution of circuit allocation times [\textbf{P4}].}
\label{fig:cat}
\end{figure}

Reusing circuits, when possible, also improves the performance of \systemname as
compared with other AS-aware Tor clients. Figure \ref{fig:cat} shows the elapsed
time between the arrival of a connection request and the allocation of a circuit
to satisfy the request. As expected, since the vanilla Tor client always uses an
existing circuit, it is significantly faster than Astoria and \systemname,
requiring under .1 seconds to allocate a circuit to over 99\% of incoming
connection requests. Within the same time constraints we see that the
\systemname Tor client is able to satisfy 60\% of its requests, while the
Astoria client can only satisfy 21\%.

Pre-emptive circuit construction yields two primary benefits. First, it is
responsible for a nearly 80\% reduction in number of relays utilized by the
AS-aware client (compared to AS-aware clients that do not do pre-emptive
construction), resulting in improved security against relay-level adversaries.
Second, it results in reduced circuit allocation times when an existing circuit
is reused. 


\subsection{Load-balance like Tor when possible (P5)} 
Load balancing is explicitly performed in two cases: (1) when constructing and
replenishing \systemname's reserve of pre-built circuits and (2) when there are
multiple safe circuits available for a connection request.

In the first case, \systemname exactly mimics the load-balancing approach
utilized by the vanilla Tor client -- \ie relays are selected in a circuit with
probability proportional to their bandwidth capacity. The second case, however,
is more nuanced. When there are multiple safe entry- and exit-relay options --
($en_1$, $ex_1$), \dots, ($en_n$, $ex_n$) -- \systemname selects the $i^{th}$
entry and exit-relay combination with probability $Pr_i$, where:

\begin{equation}
Pr_i = \frac{BW_{en_i} \times BW_{ex_i}}{\sum_{j=1}^n BW_{en_j} \times BW_{ex_j}}
\end{equation}

Here $BW_{en_j}$ and $BW_{ex_j}$ are the advertised bandwidths of the entry- and
exit-relay associated with the $j^{th}$ safe relay combination. This weighting
of combinations works to ensure that each entry- and exit-relay is selected with
the probability proportional to its advertised bandwidth (when only considering
safe relay options).

\begin{figure}[t]
\includegraphics[trim=0cm 0.1cm 0cm 2.25cm, clip=true,width=.495\textwidth]
{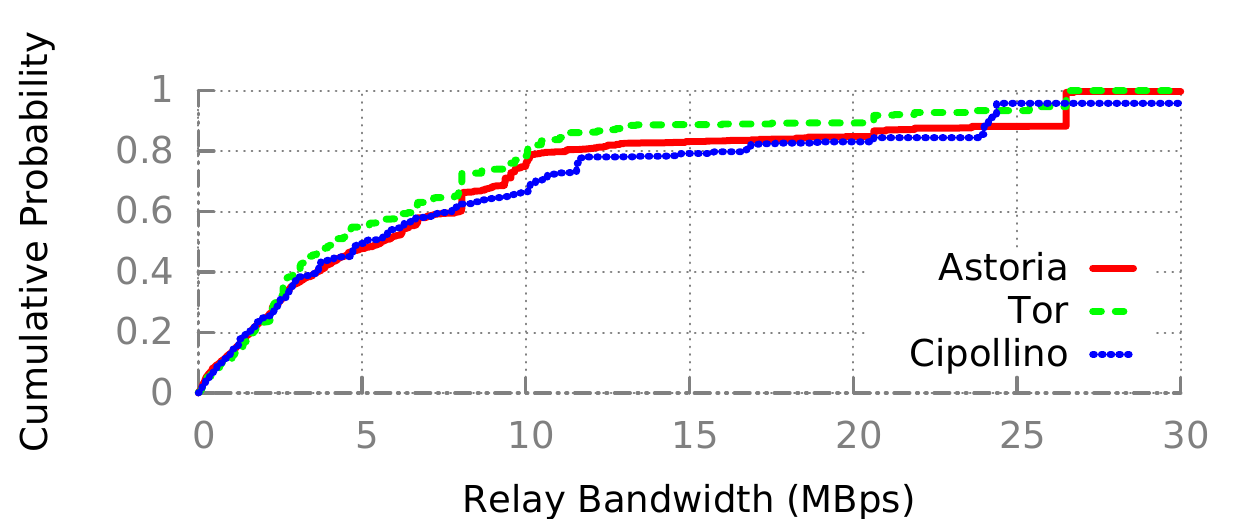}
\caption{Distributions of the bandwidths of the relays selected by each Tor
client [\textbf{P5}].}
\label{fig:load-balancing}
\end{figure}

Figure \ref{fig:load-balancing} compares the effect of the load-balancing
approaches used by the vanilla Tor client, Astoria, and \systemname. We find
that they are all able to effectively ensure that relays do not get overloaded.
Further, \systemname does not perform any worse than Astoria, despite its reuse
of existing safe circuits.

%

\subsection{Putting it all together}

In this section we describe the complete architecture of \systemname. Finally,
we complete our evaluation of the security and performance of the complete
\systemname client.

\begin{figure}[t!]
\includegraphics[trim=0cm 0cm 0cm 0cm, clip=true,width=.495\textwidth]
{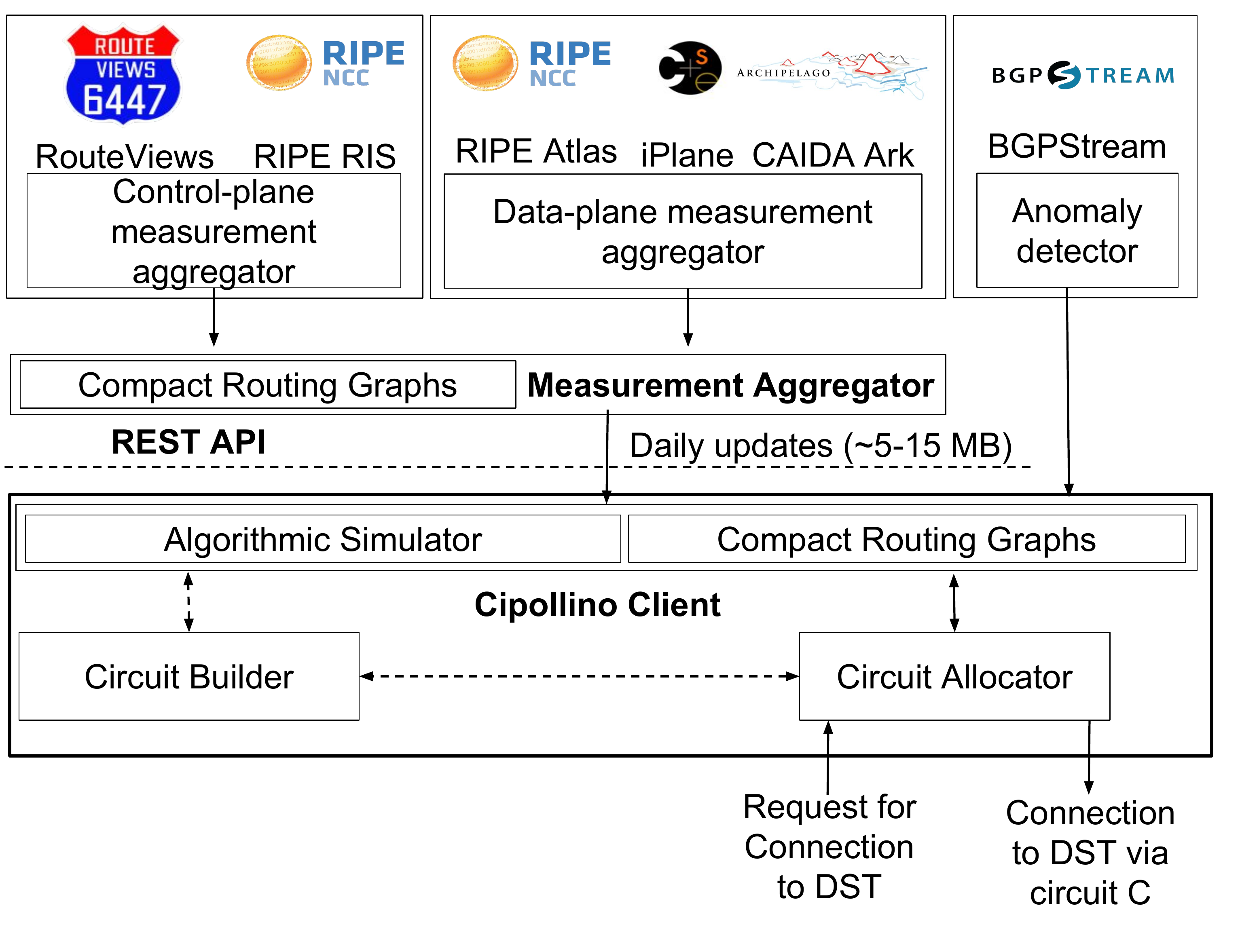}
\caption{Architecture of the \systemname Tor client.}
\label{fig:architecture}
\end{figure}

\myparab{\systemname architecture.}
\systemname consists of three main components: (1) an AS-level path aggregation
toolkit (PathCache), (2) a circuit allocator, and (3) a circuit builder. The
interaction between each of these components is illustrated in Figure
\ref{fig:architecture}.

The \systemname client maintains a compact local repository of destination-based
routing graphs. These are updated by the PathCache servers on a daily (or,
configurable) basis. The PathCache path-stitching algorithms are used on these
graphs to identify ASes that are in a position to observe traffic flowing
between a given source and destination AS.

When the \systemname client receives a request for a connection to a destination
IP and port, the circuit allocator uses the PathCache stitching algorithms and
graphs to identify if there are any pre-built circuits that are not vulnerable
to traffic correlation attacks by ASes. If exactly one of the \emph{safe}
circuits is able to serve the requested IP and port of the destination, then
the circuit is used to satisfy the connection request. If there are multiple
such circuits, then one of them is chosen in accordance with our load-balancing
scheme described in the previous section.

In the event that none of the pre-built circuits is able to satisfy the
connection, the circuit builder constructs a circuit specifically for the
requested connection. The constructed circuit performs also relay selection in a
way that achieves load-balancing across all relays in the Tor network.
Additionally, the circuit builder also handles the worst-case scenario -- when
there are no safe circuits that may be built. In this case, the circuit builder
borrows the linear program proposed by the Astoria Tor client to ensure that no
single adversary is able to de-anonymize a large number of circuits.

\begin{figure}[t!]
\includegraphics[trim=0cm 0cm 0cm 2.25cm, clip=true,width=.495\textwidth]
{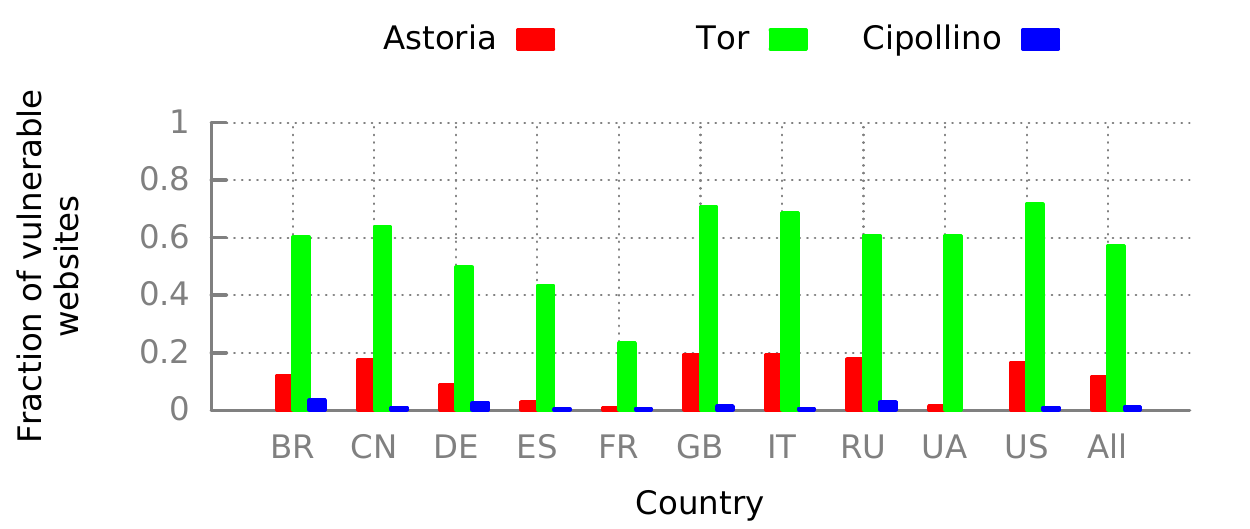}
\caption{Security of \systemname and other clients against AS-level adversaries.}
\label{fig:cipollino-security}
\end{figure}

\myparab{\systemname security against AS-level adversaries.}
We compare the security of the circuits constructed by \systemname, Astoria, and
the vanilla Tor client while performing 200 page-loads performed in each of ten
different client locations (same settings as \textbf{M1}). The results are shown
in Figure \ref{fig:cipollino-security}. From these results we see that the
\systemname client circuits provide more security against AS-level
traffic-correlation adversaries. Only 1.4\% of all webpages loaded by the
\systemname client utilized a vulnerable circuit, when compared to 11\% and 57\%
for the Astoria and vanilla Tor clients, respectively.

\begin{figure}[t!]
\includegraphics[trim=0cm 0cm 0cm 2.25cm, clip=true,width=.495\textwidth]
{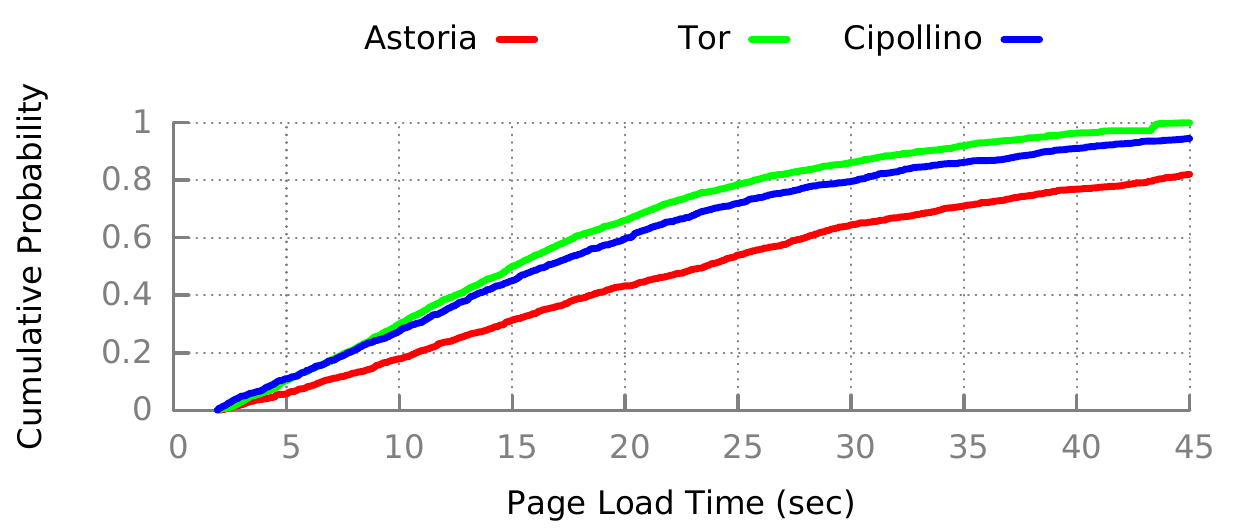}
\caption{Distribution of page-load times}
\label{fig:plt}
\end{figure}

\myparab{\systemname page-load times.}
To give a complete picture of the performance of the \systemname client we
consider the time required to load a complete web-page (including third-party
content). Figure \ref{fig:plt} shows the cumulative distribution of page-load
times of 2000 web-pages in ten client locations for the \systemname, Astoria,
and Tor clients. We find that the time required for loading pages using the
\systemname and Tor client are quite closely matched with the median page-load
time differing by only 1.6 seconds, while the Astoria Tor client is nearly 7
seconds slower. 

%% file: conclusions.tex
\section{Conclusions} \label{sec:conclusions}

In this paper we analyzed the threat faced by Tor clients from AS-level
adversaries from a current and historical perspective. We found that the current
threat is high, with around 30\% of all Tor circuits created in our experiments
remaining vulnerable to de-anonymization by AS-level correlation attacks,
regardless of whether the Tor client is used for web browsing or other
applications. Further, our historical analysis points to a fundamental problem
with the Tor network -- the lack of growth of AS-level diversity. Without
specific efforts from the Tor project to increase diversity of relays or
incorporate AS-awareness in the Tor client, our study shows that the threat is
bound to increase.

Our survey of previous work identified five common pitfalls associated with the
design and construction of AS-aware Tor clients. We show how each of these
pitfalls results in high under-estimation of the threat from AS-level
adversaries, or increased vulnerability to active (AS-level) and passive
(relay-level) adversaries, or poor performance characteristics.

We find that our AS-aware Tor client -- \systemname, designed specifically to
address these pitfalls improves the current state-of-the-art by achieving better
security against network-level adversaries. Specifically, by using a data- and
control-plane measurement infrastructure whenever possible, \systemname reduces
the fraction of vulnerable webpage loads from 57\% (vanilla Tor) and 11\%
(Astoria) to 1.4\%. Additionally, by incorporating the concept of circuit
pre-building and circuit re-use, the \systemname client significantly reduces
the threat faced from malicious relays. As a consequence of circuit pre-building
and re-use, the \systemname client is also able achieve performance
characteristics comparable with the vanilla Tor client.